\documentclass[%
 reprint,
 amsmath,amssymb,
 aps,
]{revtex4-2}

\usepackage{graphicx}
\usepackage{dcolumn}
\usepackage{bm}

\usepackage{comment}  

\usepackage{xcolor} 

\begin{document}
\preprint{APS/123-QED}

\title{Anisotropy in a wire medium resulting from the rectangularity of a unit cell}


\author{Denis Sakhno}
\email{denis.sakhno@metalab.ifmo.ru}

\author{Rustam Balafendiev}
\altaffiliation[Also at ]{Science Institute, University of Iceland, 107 Reykjavik, Iceland}

\author{Pavel A. Belov}%

\affiliation{School of Physics and Engineering, ITMO University, Kronverksky Pr. 49, 197101, St. Petersburg, Russia}


\begin{abstract}
The study is focused on the dispersion properties of a wire medium formed by a rectangular lattice of parallel wires at the frequencies close to its plasma frequency. While the effective medium theory predicts isotropic behaviour of transverse magnetic (TM) waves in the structure, numerical simulations reveal noticeable anisotropic properties. This anisotropy is dependent on the lattice rectangularity and reaches over 6\% and over 75\% along and across the wires respectively for thick wires with the radii about 20 times smaller than the smallest period.
This conclusion is confirmed by line-of-current approximation theory.
The revealed anisotropy effect is observed when the wavelength at the plasma frequency is comparable to the period of the structure. The effect vanishes in the case of extremely thin wires. 
A dispersion relation for TM waves in the vicinity of the $\Gamma$-point was obtained in a closed form. This provides an analytical description of the anisotropy effect. 

\end{abstract}

\maketitle


\section{Introduction}

Metamaterials refer to artificially created media engineered to have some particular properties, which do not appear to be observed in natural materials \cite{engheta2006metamaterials,kadic20193d, alu2024metamaterials, sakoda2019electromagnetic, engheta2023four}. 
The variety of their possible applications has been recently raising the interest of researchers to their study \cite{jahani2016all, padilla2022imaging, sun2018overview, krushynska2023emerging, poddubny2013hyperbolic}.
Wire media are a class of metamaterials composed of conducting wires periodically arranged in a host material or in a free space \cite{simovski2012wire}. Wire media feature strong spatial dispersion, even at low frequencies\cite{belov2003strong,simovski2004}, thus providing subwavelength imaging \cite{belov2006resolution,Palikaras2010} and
radiation control \cite{fernandes2012cherenkov, silveirinha2012radiation} among other manipulations of electromagnetic fields, far from common \cite{simovski2012wire}.

\begin{figure}[h]
    \begin{minipage}{0.9\linewidth}
        \center{
            \includegraphics[width=0.9\textwidth]{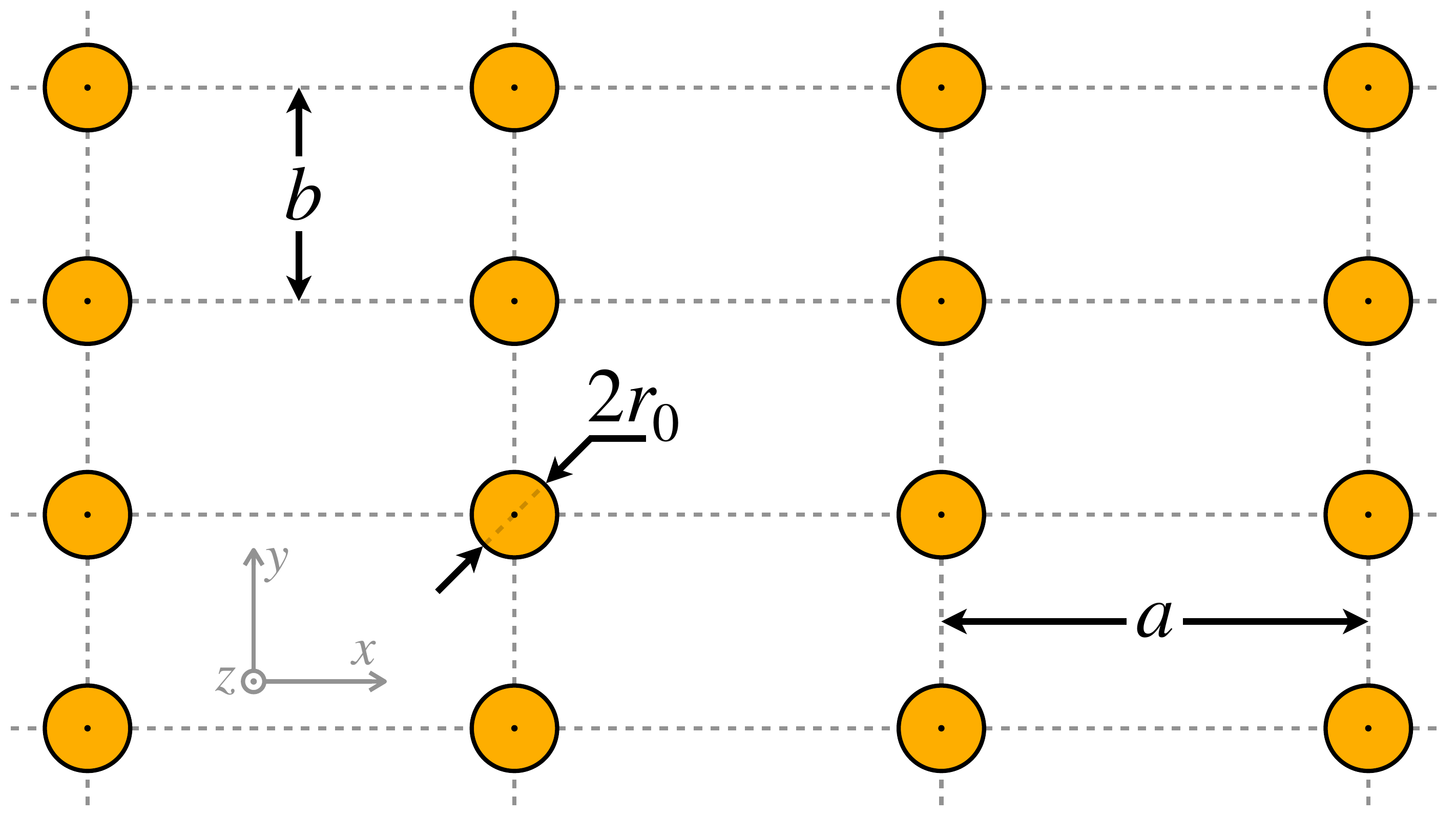}
        }
    \end{minipage}
    \caption{
    Geometry of a simple wire metamaterial formed by a rectangular lattice $a\times b$ of parallel wires of the radii equal to $r_0$.
    } \label{fig:swm} 
\end{figure}

A simple wire medium is formed by parallel metallic wires arranged periodically in a perpendicular plane. Most of the research on this  metamaterial \cite{belov2002dispersion, belov2003strong, simovski2004, belov2006resolution, Palikaras2010, fernandes2012cherenkov, silveirinha2012radiation, kowitt2023tunable} has been focused on rectangular/square periodicity, that is, an arrangement of the wires of radii $r_0$ at the nodes of a rectangular lattice having the period $a$ in the $x$-direction and the period $b$ in the $y$-direction. In the general case $a$ is not equal to $b$. The geometry of a simple wire medium formed by a rectangular lattice is shown in Fig. \ref{fig:swm}.

Square-lattice-based metamaterial ($a=b$) has been most amply studied in scholarly literature \cite{pendry1996extremely, pendry1998low}  while the rectangular configuration of the metamaterial ($a\neq b$) has rarely been included in its scope except for the recent work \cite{kowitt2023tunable}. It is, however, noteworthy that the authors of \cite{belov2002dispersion,belov2003strong} discussed the rectangular lattice of wires and presented an analytical theory of dispersion for the medium. It is based on the above theory that the present study analyses simple wire media dispersion properties.

The currently re-emerging interest to the wire medium with a rectangular lattice results from the prospects of using this metameterial for reconfigurable microwave cavities in the search of the dark matter \cite{millar2023alpha}. While wire medium filled cavities have already been analytically investigated for the case of a square lattice \cite{balafendiev2022resonator}, this study has not yet been extended to the case that relies on rectangular lattices for tuning, which is partly due to the anisotropy emerging in such systems.

\section{Anisotropy effect observed}
\begin{figure}[h]
    \begin{minipage}{1\linewidth}
        \center{
            \includegraphics[width=0.8\textwidth]{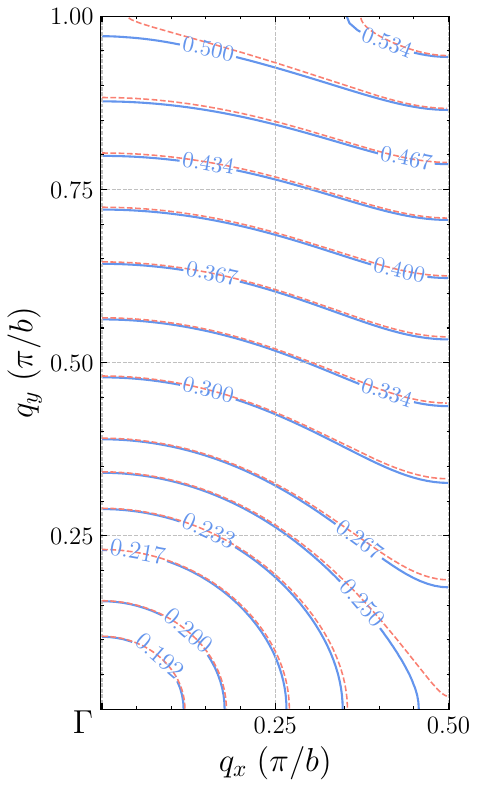}
        }
    \end{minipage}
    \caption{
    Isofrequency contours obtained numerically via COMSOL (solid lines) and analytically by solving Eq. (\ref{eq:disp_eq}) (thin dashed lines) for a metamaterial with a rectangular lattice ($a=2b$). The ratio of the minimal period $b$ to the radii of perfectly conducting wires is $b/r_0=20$. Values on iso-contours correspond to $\omega b/(2\pi c)$. Plasma frequency for the metamaterial is $\omega_p b/(2\pi c)\approx 0.185$.
    } \label{fig:isofreq_comsol_2x1} 
\end{figure}

We consider a simple wire medium formed by a rectangular lattice with an $a \times b$ unit cell, where $b$ is assumed to be half as large as $a$. The wires are assumed to be perfectly electrically conducting (PEC) and having the radii $r_0=b/20$. We have performed a numerical simulation by applying the periodic boundary conditions corresponding to the wave vector $\vec q=(q_x, q_y, q_z)^T$ for a unit cell in COMSOL Multiphysics \cite{comsol}. The isofrequency contours obtained for $q_z=0$ are plotted in Fig. \ref{fig:isofreq_comsol_2x1} with solid lines.


Figure \ref{fig:isofreq_comsol_2x1} shows that the contours cross the  $q_x$ and $q_y$ axes in the points having different coordinates $d_x$ and $d_y$ ($d_x\neq d_y$). In other words, the contours arise from the $\Gamma$-point and are elliptical before the contours reach the edges of the Brillouin zone at $q_x=\pi/a=\pi/2b$.
For the obtained contours the ratio of the ellipse's semi-axes $d_x/d_y$ (a contour's \textit{ellipticity}) tends to be $\sim 1.13$ in the vicinity of the $\Gamma$-point ($\omega=\omega_p + \delta$, $\delta\rightarrow 0$).

\textit{The electromagnetic anisotropy} for the simple wire medium resulting from the periods of the rectangular lattice changing does not appear to have been ever discussed earlier. However, an analogous \cite{nicolas1998analogy} acoustic task for a two-dimensional array of rigid cylinders was reported in \cite{torrent2008anisotropic}, where an anisotropy for an acoustic wave was analytically and numerically shown in the long wavelength limit.

The observed contour \textit{ellipticity} can not be described by the dispersion equation derived in \cite{belov2002dispersion,belov2003strong}:
\begin{equation}
    q^2=q_x^2+q_y^2+q_z^2=k^2-k_p^2,
    \label{eq:taylor_circle}
\end{equation}
that predicts circular contours regardless of the periods ratio $a/b$. 
Equation (\ref{eq:taylor_circle}) was obtained in \cite{belov2002dispersion} by applying the assumption of a small wavenumber within the metamaterial $q$ ($qa,qb\ll \pi$) and a small vacuum wavenumber $k$ ($ka,kb\ll \pi$) in \textit{the original (complete) dispersion equation}:
\begin{eqnarray}
    F&&(\boldsymbol{q}, k, a,b,r_0)= 0,
    \label{eq:disp_eq}
\end{eqnarray}
where
\begin{eqnarray}
    F&&(\boldsymbol{q}, k, a,b,r_0)=
    \frac{1}{\pi}\ln{\frac{b}{2\pi r_0}} +
    \frac{1}{k_x^{(0)}b}\;
    \frac{\sin{k_x^{(0)}a}}{\cos{k_x^{(0)}a} - \cos{q_xa}}
    \nonumber\\
    && +
    \sum\limits_{n\neq0}\left( \frac{1}{k_x^{(n)}b} \;
    \frac{\sin{k_x^{(n)}a}}{\cos{k_x^{(n)}a} - \cos{q_xa}} - \frac{1}{2\pi|n|} \right)=0.
    \label{eq:func}
\end{eqnarray}
In Eq. (\ref{eq:func}) $k_x^{(n)}=-j\sqrt{(2\pi n/b+q_y)^2+q_z^2-k^2}$ and $k=\omega/c$. From now on we will refer to Eq. (\ref{eq:taylor_circle}) as to \textit{a low-$k\&q$ model}. The plasma wavenumber $k_p$ in this model is expressed as
\begin{equation}
    k_p^2=\frac{2\pi / (ab)}{\log{\frac{b}{2\pi r_0}} + \sum\limits_{n=1}^{+\infty} \frac{\coth{\frac{\pi n a}{b} - 1}}{n} + \frac{\pi a}{6b}}.
    \label{eq:k0_circle}
\end{equation}
Equation (\ref{eq:k0_circle}) provides an adequate estimation of the plasma frequency even for the thick wires ($b/r_0=20$) we used in the simulation above ($\omega_pb/2\pi c \approx 0.190$ versus $0.185$ found in COMSOL).

The red dashed lines in Fig. \ref{fig:isofreq_comsol_2x1} refer to the analytical isofrequency contours calculated using the dispersion Eq. (\ref{eq:disp_eq}). The proper correspondence between the numerical and analytical results (up to the moment when the elliptical contour reaches the edges of the Brillouin zone) confirms the effect of the contour ellipticity and confirms the possibility of using Eq. (\ref{eq:disp_eq}) to explain the elliptical shape of contours. 

We have also verified that a direct numerical solution of the transcendental Eq. (\ref{eq:disp_eq}) matches the results of the computational simulation in COMSOL for the wire media with other ratios $a/b\leq10$ and other wire radii $b/r_0\geq20$. Based on these results we can suggest that Eq. (\ref{eq:disp_eq}) can be an adequate substitute for the eigenmode solver in COMSOL for a simple wire medium composed of thin wires ($b/r_0 \geq 20$) in the vicinity of the $\Gamma$-point.

\section{Analytical model of anisotropy}

In \cite{belov2002dispersion} the low-$k\&q$ model 
(Eq. (\ref{eq:taylor_circle})) was derived from Eq. (\ref{eq:disp_eq}) with the assumption of the parameters being small: (1) $\boldsymbol{q}$ ($qa,qb\ll \pi$) and (2) $k$ ($ka,kb\ll \pi$). However, this derivation is no longer applicable if $k$ is comparable with $\pi/a$ or $\pi/b$. For the example provided in Fig. \ref{fig:isofreq_comsol_2x1} ($a=2b$, $b/r_0=20$), the wavenumber $k$ for the first eigenmode of the metamaterial is close to the plasma wavenumber $k_p$, which is equal to $0.37\pi/b$, a value comparable to $\pi/b$. The low-$k\&q$ model not being applicable in this case, a new model has to be developed.

We propose to modify the aforementioned model 
by keeping the assumption about the small $q$ ($qa \ll \pi$ and $qb \ll \pi$) and removing the condition of a small $k$, thus obtaining 
\textit{the low-$q$ model}.
The expansion of the Eq. (\ref{eq:disp_eq}) into Taylor series up to the second order of small parameters ($q_ia$ and $q_ib$, $i=x,y,z$) results in a more complicated dispersion equation near the $\Gamma$-point:
\begin{eqnarray}
    A(k,a,b) q_x^2 + B(k,a,b) q_y^2 + C&&(k,a,b) q_z^2
    \nonumber \\
    &&=F_0(k,a,b,r_0).
    \label{eq:disp_ellipse_0}
\end{eqnarray}
Terms of the first order and cross-terms are equal to zero due to the symmetry of the metamaterial geometry.
Equation (\ref{eq:disp_ellipse_0}) divided by function $F_0$  results in a classical form of a quadric surface:
\begin{equation}
    \frac{q_x^2}{d_x^2} + \frac{q_y^2}{d_y^2} + \frac{q_z^2}{d_z^2} = 1.
    \label{eq:disp_ellipse_1}
\end{equation}
The coefficients of $q_i^2$ ($i=x,y,z$) are the inverse squares of the ellipsoid semi-axes lengths: $d_x=\sqrt{F_0/A}$, $d_y=\sqrt{F_0/B}$ and $d_z=\sqrt{F_0/C}$.

The coefficients $F_0$, $A$, $B$ and $C$ in Eq. (\ref{eq:disp_ellipse_0}) are provided by the following expressions in a closed form:
\begin{eqnarray}
    F_0= && 
    \frac{1}{\pi} \ln{ \frac{b}{2\pi r_0} }
    - \frac{1}{kb} \cot{\left( \frac{ka}{2} \right)}
    \label{eq:f0} \\
    && +
    \frac{1}{\pi}
    \sum\limits_{n=1}^\infty 
    \left[
    \frac{2\pi \coth{ \left( \frac{a}{2b} \psi_n(k) \right) } }
    { \psi_n(k) }
    - \frac{1}{n}
    \right],
    \nonumber \\
    A= &&
    \frac{a^2}{4} \left[
    \frac{1}{kb} \sin^{-2}{\left(\frac{ka}{2}\right)} \cot{\left(\frac{ka}{2}\right)}\right. \label{eq:a} \\
    &&
    \left. 
    + 2 
    \sum\limits_{n=1}^\infty 
    \frac{1}{ \psi_n(k) }
    \frac{ \cosh{ \left( \frac{a}{2b} \psi_n(k) \right) } }{\sinh^3{ \left( \frac{a}{2b} \psi_n(k) \right) }}
    \right],
    \nonumber \\
    B= &&
    \frac{a}{2k^2b}\cot{\left(\frac{ka}{2}\right)}
    \left[ \frac{1}{ka} + \frac{1}{\sin{ ka }} \right]  \label{eq:b} \\
    &&+ 
    \sum\limits_{n=1}^\infty 
    \frac{b^2}{ \psi_n^3(k) }\left(
    1
    -\frac{12(\pi n)^2 }{ \psi_n^2(k) }
    \right)
    \coth { \left( \frac{a}{2b} \psi_n(k) \right) } 
    \nonumber \\
    && + 
    \sum\limits_{n=1}^\infty 
    \frac{ ab/2 }{ \psi_n^2(k) }
    \left(
    1
    - \frac{ 12 (\pi n)^2 }{ \psi_n^2(k) }
    \right)
    \sinh^{-2}{ \left( \frac{a}{2b} \psi_n(k) \right) }
    \nonumber\\
    &&-
    \sum\limits_{n=1}^\infty 
    \frac{2a^2  (\pi n)^2}{ \psi_n^3(k) } 
    \frac{ \cosh{ \left( \frac{a}{2b} \psi_n(k) \right) } }{\sinh^3{ \left( \frac{a}{2b} \psi_n(k) \right) }},
    \nonumber
\end{eqnarray}
\begin{eqnarray}
    C= &&
    \frac{a}{2k^2b}\cot{\left(\frac{ka}{2}\right)}
    \left[ \frac{1}{ka} + \frac{1}{\sin{ ka }} \right] \label{eq:c} \\
    && +
    \sum\limits_{n=1}^\infty \frac{ b^2 }{ \psi_n^3(k) }
    \coth { \left( \frac{a}{2b} \psi_n(k) \right) } 
    \nonumber \\
    &&+ 
    \sum\limits_{n=1}^\infty \frac{ ab/2 }{ \psi_n^2(k) } 
    \sinh^{-2}{ \left( \frac{a}{2b} \psi_n(k) \right) },
    \nonumber
\end{eqnarray}
where $\psi_n(k)=\sqrt{(2\pi n)^2 - (kb)^2}$.

Since the system under consideration does not change with the substitutions $q_x \leftrightarrow q_y$ and $a \leftrightarrow b$  performed simultaneously, the following conditions are required for the coefficients:
\begin{equation}
    \begin{matrix}
        A(k,a,b)&=&B(k,b,a)\\
        C(k,a,b)&=&C(k,b,a)\\
        F_0(k,a,b,r_0)&=&F_0(k,b,a,r_0).
    \end{matrix}
    \label{eq:symmetry}
\end{equation}
These properties are quite hard to see from the expressions (\ref{eq:f0}-\ref{eq:c}), but we have checked them numerically.

Each of the functions $F_0$, $A$, $B$ or $C$ can be calculated for an arbitrary $k$ close to $k_p$ and, hence, isofrequency contours in the low-$q$ model can be plotted without solving Eq. (\ref{eq:disp_ellipse_0}) in each point of the mesh $(q_x, q_y, q_z)^T$. On the other hand, to obtain the plasma wavenumber $k_p$ we have to find the first root of the transcendental equation:
\begin{equation}
    F_0(k_p,a,b,r_0) = F(k_p, \boldsymbol{0}, a, b, r_0) = 0.
    \label{eq:k0_ellipse}
\end{equation}

\begin{figure}[h]
    \begin{minipage}{1\linewidth}
        \center{
            \includegraphics[width=1\textwidth]{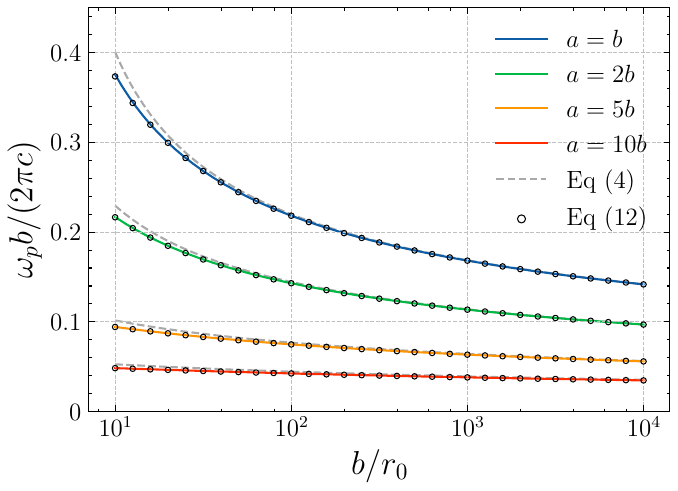}
        }
    \end{minipage}
    \caption{
    Plasma frequency (obtained in different ways) dependence on $b/r_0$ ratios for different $a/b$ values. Solid lines: plasma frequencies calculated via COMSOL, dashed lines: those obtained using Eq. (\ref{eq:k0_circle}). Circle markers were derived using Eq. (\ref{eq:k0_ellipse}).
    } \label{fig:plasma} 
\end{figure}

We also compared the performance of Eqs. (\ref{eq:k0_circle}) and (\ref{eq:k0_ellipse}) to evaluate the plasma frequency against the numerical results. A plot of plasma frequency dependencies on $b/r_0$ ratio for different $a/b$ relations is shown in Fig. \ref{fig:plasma}.
The plasma frequency was calculated by three different means: (i) using COMSOL (the actual plasma frequency), the result is represented by solid lines in the plot; (ii) using Eq. (\ref{eq:k0_circle}), the result is shown in dashed lines; (iii) using Eq. (\ref{eq:k0_ellipse}), the result is depicted in circular markers. Equation (\ref{eq:k0_ellipse}) was shown to perform perfectly: the results of numerical modeling match the analytical results. Hereafter the value of $b$ is maintained fixed to make a unified normalization of all the results.

\begin{figure*}
    \begin{minipage}{1\linewidth}
        \center{
            \includegraphics[width=1\textwidth]{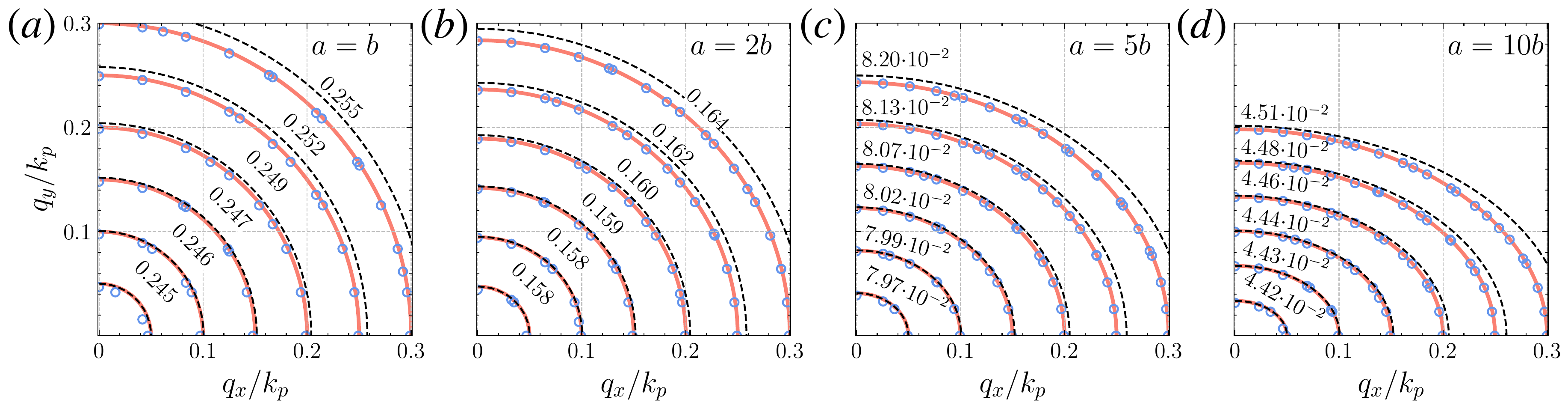}
        }
    \end{minipage}
    \caption{
    Isofrequency contours near the $\Gamma$-point calculated using Eq. (\ref{eq:disp_eq}) (solid lines) and the low-$q$ model (Eq. (\ref{eq:disp_ellipse_0})) (dashed contours) for different geometries of the metamaterial: (a) $a=b$, (b) $a=2b$, (c) $a=5b$ and (d) $a=10b$. For all cases $b/r_0=50$ was fixed. The plot axes were normalised by the corresponding $k_p$: (a) $0.489\pi/b$, (b) $0.315\pi/b$, (c) $0.159\pi/b$ and (d) $0.088\pi/b$. Circle markers indicate the contour points obtained by extracting the contours from numerical simulations.
    }
    \label{fig:isofreqs_taylor}
\end{figure*}

The greater are radii of wires $r_0$, the greater is the deviation of the frequency estimated by Eq. (\ref{eq:k0_circle}) from the actual value obtained in numerical calculations. Table \ref{tab:plasma} provides the relative errors of the plasma frequency estimations by Eqs. (\ref{eq:k0_circle}) and (\ref{eq:k0_ellipse}) 
\begin{table}[h]
    \caption{\label{tab:plasma}
    Relative error $\xi$ of the plasma frequency estimations obtained from Eqs. (\ref{eq:k0_circle}) and (\ref{eq:k0_ellipse}) for different $a/b$ ratios and the same wires radii $b/r_0=10$.}
    \begin{ruledtabular}
        \begin{tabular}{c|ccc}
            $a/b$&
            $\omega_p^{\text{(num)}}b/(2\pi a)$ &
            $\xi^\text{(Eq. \ref{eq:k0_circle})}$, \%&
            $\xi^\text{(Eq. \ref{eq:k0_ellipse})}$, \%\\
            \colrule
            1 & 0.3753 & 6.728 & 0.535 \\
            2 & 0.2168 & 5.841 & 0.128 \\
            5 & 0.0944 & 7.664 & 0.016 \\
            10 & 0.0486 & 8.804 & 0.005 \\
        \end{tabular}
    \end{ruledtabular}
\end{table}
for the smallest $b/r_0$ ratio from Fig. \ref{fig:plasma} ($b/r_0=10$) and different $a/b$ ratios. It is noteworthy that Eq. (\ref{eq:k0_ellipse}) provides perfect accuracy of the estimation even for \textit{thick} wires.

\section{Verification of the analytical model}

We have compared the metamaterial's anisotropy effects obtained numerically with those described by the low-$q$ model (\ref{eq:disp_ellipse_1}) to verify the ability of analytical model to account for these effects. Figure \ref{fig:isofreqs_taylor} shows the comparison of the three means of calculating isofrequency contours in the $xy$-plane. The first one is solving the original dispersion Eq. (\ref{eq:disp_eq}) (solid lines in the figure), the second one uses the low-$q$ model (Eq. (\ref{eq:disp_ellipse_1}), plotted by dashed lines). The third calculation was made using COMSOL Multiphysics full-wave simulation \cite{comsol}, and these numerical results are represented by circle markers in the Figure. The plots are provided in normalized axes. 

\begin{figure}
    \begin{minipage}{1\linewidth}
        \center{
            \includegraphics[width=1\textwidth]{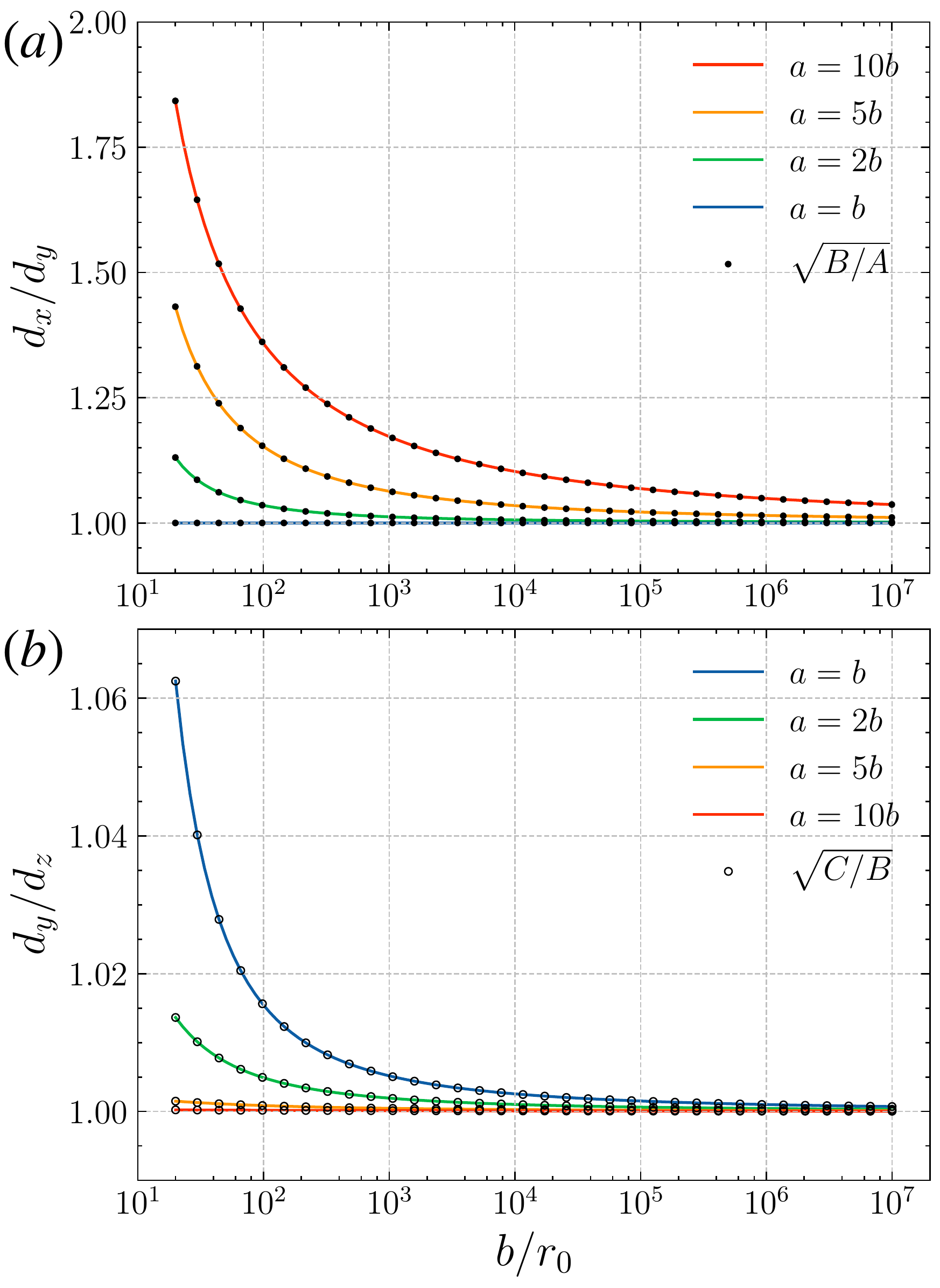}
        }
    \end{minipage}
    \caption{
    Dependencies of the ellipticity on $b/r_0$ ratio for different $a/b$ values. (a) Solid lines plotted by calculating of $q_x/q_y$, where $q_x$ and $q_y$ are the solutions of Eq. (\ref{eq:disp_eq}) for fixed $k$ near $k_p$. Black dots obtained as $\sqrt{B/A}$, where $A(k,a,b)$ and $B(k,a,b)$ are the functions defined in Eqs. (\ref{eq:a}) and (\ref{eq:b}). (b) The same comparision between ellipticities in $yz$-plane obtained via Eq. (\ref{eq:disp_eq}) (solid lines) and by calculating $C(k,a,b)$ and $B(k,a,b)$ functions (\ref{eq:b}) and (\ref{eq:c}) (circle markers).
    } \label{fig:ellipticity} 
\end{figure}

The numerical results perfectly match the analytics provided by Eq. (\ref{eq:disp_eq}). Slight inconsistencies of the markers compared to the solid lines are explained by different mesh resolutions that were used for numerical and analytical calculations near the $\Gamma$-point and by the approximation used for numerical data to obtain contours. 

Figure \ref{fig:isofreqs_taylor} shows that the low-$q$ model diverges from the Eq. (\ref{eq:disp_eq}) with the increase in the frequency, while the matching is nearly perfect in the vicinity of the $\Gamma$-point for every geometry of the metamaterial. Therefore, the proposed model perfectly describes the ellipticity of contours at the frequencies slightly greater than the plasma frequency.

To demonstrate the ability of the low-$q$ model (\ref{eq:disp_ellipse_0}) to predict the ellipticity of contours for different configurations of the metamaterial we performed the calculation of $d_x/d_y$ and $d_y/d_z$ ratios for a wide range of geometries, see Fig. \ref{fig:ellipticity}. The circle markers in Fig. \ref{fig:ellipticity} were calculated (a) as $\sqrt{B / A}$ using expressions (\ref{eq:a}-\ref{eq:b}) and (b) as $\sqrt{C / B}$ by calculating the functions (\ref{eq:b}-\ref{eq:c}). The solid lines in both subfigures were obtained by solving Eq. (\ref{eq:disp_eq}) for fixed $k$ three times: (i) for $q_x$ with fixed $q_y=q_z=0$, (ii) for $q_y$ with fixed $q_x=q_z=0$ and (iii) for $q_z$ with fixed $q_x=q_y=0$ at $k$ near $k_p$. These two calculations of ellipticity have been shown to perfectly correspond to each other in both planes: $xy$ and $yz$.

It is important to note, that for extremely thin wires ($b/r_0\rightarrow\infty$) the shape of isofrequency contours tends to be circular. This is explained by the plasma frequency decreasing with the decrease of wire radii (Fig. \ref{fig:plasma}), which makes the low-$k\&q$ model applicable because of $ka$ and $kb$ becoming small enough. The higher the $a/b$ ratios, the more difficult it is to achieve a small $ka$ value, hence, the ellipticity for these geometries tends to $1$ for very high $b/r_0$ values. 

The results shown in Fig. \ref{fig:ellipticity}(b) prove to demonstrate the ellipticity even for the metamaterials with a square lattice, however, in $yz$-plane only. The maximum value for the $d_y/d_z$ ratio is equal to $\sim 1.06$. The ellipticity in $xy$-plane (Fig. \ref{fig:ellipticity}(a)) achieves the maximal value of $\sim 1.80$ in our simulations for the case of a rectangular lattice with the $a/b$ ratio equal to $10$.

\section{Conclusion}
The present study was focused on a simple wire medium formed by a rectangular lattice of parallel wires. The research revealed the anisotropy of TM waves near the $\Gamma$-point. To describe the anisotropy we proposed and tested the low-$q$ model that is applicable for anisotropy prediction near the Brillouin zone center.

The results of this work are valuable in the context of the axion search, since simple wire medium has recently served as a basis for axion haloscopes \cite{millar2023alpha, kowitt2023tunable}.
An accurate prediction of the extent of anisotropy in a given rectangular lattice is crucial for the estimation of the tuning ranges for the projects on the dark matter search relying on rectangular wire media for tuning.
The tunability of haloscopes is crucial because of the requirement to cover the frequency range not yet studied. Hence, the possibility to employ a rectangular-lattice-based simple wire medium with a precise understanding of anisotropy dependence on the lattice periods paves the way to new opportunities for tuning the haloscope.




\nocite{*}

\bibliography{apssamp}

\end{document}